\documentclass[preprint,showpacs,bibnotes,amsmath,amssymb,aps,prb]{revtex4-1}
\usepackage[T1]{fontenc}
\usepackage[latin9]{inputenc}
\usepackage{amsmath}
\usepackage{esint}

\makeatletter
\usepackage{bbold}

\makeatother

\begin{document}

\title{Reexamination of the Elliott-Yafet Spin-Relaxation Mechanism}

\author{Alexander Baral}
\affiliation{Physics Department and Research Center OPTIMAS, Kaiserslautern University,
P. O. Box 3049, 67663 Kaiserslautern, Germany}

\author{Svenja Vollmar}
\thanks{Graduate School of Excellence Materials Science in Mainz, Gottlieb-Daimler-Strasse
47, 67663 Kaiserslautern, Germany}
\affiliation{Physics Department and Research Center OPTIMAS, Kaiserslautern University,
P. O. Box 3049, 67663 Kaiserslautern, Germany}

\author{Steffen Kaltenborn}
\affiliation{Physics Department and Research Center OPTIMAS, Kaiserslautern University,
P. O. Box 3049, 67663 Kaiserslautern, Germany}

\author{Hans Christian Schneider}
\email{hcsch@physik.uni-kl.de}
\affiliation{Physics Department and Research Center OPTIMAS, Kaiserslautern University,
P. O. Box 3049, 67663 Kaiserslautern, Germany}

\renewcommand{\vec}[1]{\mathbf{#1}}

\date{\today}
\begin{abstract}
We analyze spin-dependent carrier dynamics due to incoherent electron-phonon scattering, which is commonly referred to as Elliott-Yafet (EY) spin-relaxation mechanism. For this mechanism one usually distinguishes two contributions:  (1) from the electrostatic interaction together with spin-mixing in the wave functions, which is often called the Elliott contribution, and (2) the phonon-modulated spin-orbit interaction, which is often called the Yafet or Overhauser contribution. By computing the reduced electronic density matrix, we improve Yafet's original calculation, which is not valid for pronounced spin mixing as it equates the pseudo-spin polarization with the spin polarization. The important novel quantity in our calculation is a torque operator that determines the spin dynamics. The contribution (1) to this torque vanishes exactly. From this general result, we derive a modified expression for the Elliott-Yafet spin relaxation time.
\end{abstract}

\pacs{75.78.Jp, 72.25.Rb, 76.30.Pk }

\maketitle

\section{Introduction}

Spin relaxation plays a role in spin-dependent dynamics both on long and short timescales. While its contribution to (precessional) magnetization damping is usually ascribed to spin-lattice relaxation and treated phenomenologically, the analysis of ultrafast demagnetization dynamics has often been based on the microscopic concept of spin-flip processes due to electron-phonon interactions as developed for semiconductors in the 1950s.~\cite{Zutic:2004vi} Overhauser~\cite{Overhauser:1953wi} was the first to identify the modulation of the spin-orbit interaction by lattice vibrations as the microscopic mechanism for spin relaxation due to incoherent electron-phonon scattering processes. Elliott~\cite{Elliott:1954wy} argued shortly thereafter that there is an additional contribution to the spin relaxation due to the momentum-dependent spin mixing in the wave functions and that, consequently, even spin-diagonal incoherent scattering processes due to spin-diagonal interactions can contribute to spin relaxation. The subject was taken up again by Gerasimenko and Andreev,~\cite{Andreev:1959uh} and Yafet.~\cite{Yafet:1963iw} The latter calculated the spin-flip matrix element due to electron-phonon interaction (as a function of electronic momentum transfer $q$) including both the Overhauser and the Elliott contributions and showed that the first few orders in $q$ vanish due to a cancellation of Overhauser and Elliott contributions. Nowadays, this combination of Overhauser and Elliott contributions is usually called Elliott-Yafet mechanism because Yafet derived a relatively simple result for the close-to-equilibrium \emph{spin relaxation time}, which is suitable for evaluation from ab-initio input and can be used to justify an approximate relation between the spin relaxation time and the momentum relaxation time. At present, both simplified and ab-initio based expressions due to the EY mechanism are widely used, in particular for magnetization dynamics
in metals.~\cite{Koopmans:PRL2005,Steiauf.2009,Carva:2011dp,Essert.2011,Koopmans:NatMat2010}

In this paper, we present a new analysis of the spin-relaxation problem due to \emph{incoherent} electron-phonon scattering, as it was originally considered by Overhauser, Elliott, and Yafet. We derive the dynamical equation for the change of the reduced electronic spin density matrix by expressing the spin dynamics in terms of phonon assisted density matrices. This approach achieves a correct description of the dynamics of the spin \emph{vector}, as opposed to the Yafet treatment, which gets spins and pseudospins mixed up, and thus cannot correctly account for the ``amount of spin-flip'' in each scattering transition. If one correctly describes the spin vector, it becomes obvious  that the important quantity for spin dynamics is a torque matrix element, which is not present in the conventional derivation.~\cite{Yafet:1963iw} Remarkably, there is no contribution to this torque matrix element from spin-diagonal scattering mechanisms. Put differently, spin-diagonal electron-phonon scattering and spin-orbit coupling alone, which is usually referred to as the Elliott spin-relaxation mechanism, yields \emph{no spin dynamics}. Based on this observation, we derive a modified result for the EY spin relaxation time.

This paper is organized as follows. In Sec.~\ref{E-Y} we briefly review the conventional Elliott-Yafet treatment, some more recent contributions, and how the Elliott-Yafet spin relaxation mechanism has been applied in theoretical models for the demagnetization of ferromagnets. In Sec.~\ref{e-pn-dynamics} we set up the electron-phonon interaction hamiltonian, discuss long and short-range contributions, and derive the equations of motion for the spin density matrix and the spin expectation value. Section~\ref{sec:relaxation-time} is devoted to the derivation of a spin relaxation time for the special case of Kramers degenerate bands. An important ingredient for this derivation is the form of the quasi-equilibrium spin-density matrix in the presence of spin-orbit coupling, which is discussed in some detail. The conclusions are presented in Sec.~\ref{sec:conclusion}, and appendix~\ref{sec:spin-expectation} contains a short demonstration concerning the form of the spinor wavefunctions on Kramers degenerate bands.

\section{Elliott-Yafet approach\label{E-Y}}

The Elliott-Yafet approach has been reviewed often.~\cite{pikus-titkov:opt-orient,Zutic:2004vi,Haegele-Adv-SolState} Only for comparison with our calculations, we repeat here some of the main points of Yafet's derivation in Ref.~\onlinecite{Yafet:1963iw} using his notation. The objective of Yafet is to calculate the rate of a spin-flip transition for two Kramers degenerate bands including spin-orbit coupling. The Kramers degeneracy implies that for a band index $b$ we have two wave functions
\begin{align}
\psi_{b\Uparrow\vec{k}}(\vec{x}) &= \phi_{b\uparrow,\vec{k}}(\vec{x})|\uparrow\rangle +\chi_{b\downarrow,\vec{k}}(\vec{x})|\downarrow\rangle \label{psi-uparrow} \\
\psi_{b\Downarrow\vec{k}} (\vec{x}) &=\hat{K} \psi_{\vec{k},b\Uparrow} \label{psi-downarrow}
\end{align}
with the same energy $E_{b\vec{k}}$, where $\hat{K}$ is the time-reversal operator. Focusing on one band and dropping the corresponding $b$ index he then calculates the Golden-Rule transition probability $W_{\Uparrow\vec{k},\Downarrow\vec{k'}}$ for a spin-flip transition $\psi_{\Uparrow\vec{k}}\to \psi_{\Downarrow\vec{k}}$ due to the electron-phonon interaction. More precisely, Yafet calculates the dynamics of the spin polarization, which is \emph{defined} as
\begin{equation}
\label{def-D}
D_{\mathrm{spin}}=\frac{1}{V}\sum_{\vec{k}}[n_{\Uparrow\vec{k}} - n_{\Downarrow\vec{k}}],
\end{equation}
where $n_{\lambda\vec{k}}$ denote the carrier distributions with momentum $\vec{k}$ and pseudospins $\lambda=\Uparrow$, $\Downarrow$. It is clear that $D_{\text{spin}}$ cannot be a good approximation to the spin polarization for pronounced spin mixing. Then (including a factor of 2 as in Yafet's derivation) one obtains
\begin{equation}
	\frac{d}{dt}D_{\mathrm{spin}}= 2 (W_{\Uparrow,\Downarrow}- W_{\Downarrow,\Uparrow})
	=4W_{\Uparrow,\Downarrow}.
\end{equation}
Thus the change of spin polarization is determined essentially by the number of transitions (per unit time), which is obtained by 
adding the in-scattering and out-scattering Golden-Rule probabilities
\begin{equation}
	W_{\Uparrow,\Downarrow}=\frac{1}{\mathcal{V}^2}\sum_{\vec{k},\vec{k}'}\Big\{W_{\Downarrow\vec{k},\Uparrow\vec{k}'} \,n_{\Downarrow\vec{k}}[1-n_{\Uparrow\vec{k'}}]
	-W_{\Uparrow\vec{k},\Downarrow\vec{k}'} \,n_{\Uparrow\vec{k}}[1-n_{\Downarrow\vec{k}'}]\Big\},
\end{equation}
where 
\begin{equation}
	W_{\Uparrow\vec{k},\Downarrow\vec{k}'}  = \frac{2\pi}{\hbar} \big| M_{\Uparrow\vec k, \Downarrow\vec k'}^{(\lambda)}\big|^2 
	\frac{\hbar}{2MN\omega_q} \Big[ \delta (E_{\vec{k}} - E_{\vec{k}'}+\hbar\omega_{q})N_q
	 +\delta (E_{\vec{k}} - E_{\vec{k}'}-\hbar\omega_{-q})(N_{-q}+1)\Big]
\end{equation}
and $q=\left|\vec{k'}-\vec{k}\right|$. The other symbols have an obvious meaning and are defined below. To obtain a relaxation time valid for a small spin polarization, the distribution functions $n_{\lambda \vec{k}}$  are assumed to be of the quasi-equilibrium form $f(\epsilon_{\mu\vec{k}}-\mu_\lambda)$ with $\mu_\lambda$ the (pseudo)spin dependent chemical potentials, and the distributions are expanded for small $\mu_\lambda$'s. This treatment has recently been extended to ferromagnets.~\cite{Steiauf:2010ea} Finally, Yafet shows from a symmetry argument that there is a cancellation between the two contributions to the electron-phonon (e-pn) coupling matrix element in the long-wavelength limit. In this paper, we refer to these two contributions as the Elliott and Overhauser contributions, which we explain in detail below.

In our view this method has three problems. 
\begin{itemize}
\item[1)] The spin polarization $D_{\mathrm{spin}}$, defined in \eqref{def-D}, is computed as the difference of pseudospin occupation numbers as if the electrons were in pure spin states for all $\vec{k}$ values. Here, the Yafet derivation ignores spin mixing, which leads to a $\vec{k}$ dependent expectation value of $\hat{s}_z$. Importantly, the modulus of this spin expectation value may be \emph{significantly smaller} than~$\hbar/2$.
\item[2)] The calculation of spin dynamics is based on transition probabilities between pseudo-spin states or, equivalently, pseudo-spin occupation numbers~$n_{\Uparrow\vec{k}}$ and $n_{\Downarrow\vec{k}}$. Such a treatment neglects coherences between the pseudospin states, which are the off-diagonal elements of the spin density matrix
\begin{equation}
	\rho(\vec{k}) = \begin{pmatrix}
    n_{\Uparrow\vec{k}}       & \rho^{\Uparrow,\Downarrow}_{\vec{k}} \\  
     { \rho^{\Downarrow,\Uparrow}_{\vec{k}}}    & n_{\Downarrow\vec{k}}
\end{pmatrix} .
\end{equation} 
\item[3)] Yafet proves the cancellation between the Elliott and Overhauser contributions to the spin-flip matrix element $M_{\Uparrow\vec{k},\Downarrow\vec{k}'}$ for the short range part of the electron-phonon coupling matrix element. As shown by Grimaldi and Fulde~\cite{Grimaldi:1997td} there is also a long-range contribution, for which this cancellation does not hold, and which is larger than the short-range contribution in the long-wave limit. 
 \end{itemize}
 
The Yafet method is so widely accepted that it may be worthwhile to mention that ours is not the first paper to point to these problems. For instance, Ref.~\onlinecite{Steiauf:2010ea} states
\begin{quotation}
 Obviously, $m_{\vec{k}}^{\lambda}$ [the magnetic moment of state $(\lambda,\vec{k})$] is different for different $k$-vectors, but it is about $\pm\mu_B$ as long as the spin mixing is small. Yafet neglects the $k$-dependence of $m_{\vec{k}}^{\lambda}$ and assumes a constant magnetic moment $m$ or $-m$ for all dominant spin-up or all dominant spin-down states, respectively, which is certainly problematic in systems with spin ``hot spots,'' i.e., with regions in the Brillouin zone where the spin mixing is very large.
 \end{quotation}
Also, for semiconductors, Yu et al.~\cite{Yu:2005tp} have used spin projection operators to avoid the first problem in their calculation of the EY spin-relaxation time, but missed the Overhauser contribution, as they only include spin-diagonal scattering processes. 
 
Going beyond spin-relaxation times, there are microscopic approaches to spin-dependent carrier dynamics that compute the reduced density matrix and thus avoid using the pseudospin polarization,~\cite{Wu-Metiu,Glazov-Ivchenko,Krauss-Bratschitsch,Shen-Wu} see Ref.~\onlinecite{Wu:2010jf} for a review. Numerically solving for the full dynamical spin density matrix also yields Dyakonov-Perel and Bir-Aronov-Pikus contributions spin relaxation, but is very CPU-time intensive and difficult because the microscopic carrier dynamics has to be computed to obtain spin relaxation times that are often orders of magnitude longer than typical scattering times, which have to be resolved in the numerics. Numerically calculating the spin density matrix does not yield explicit expressions for spin relaxation times, which is one of the goals of our paper. Further, our calculation gives a more transparent description of the ensemble spin dynamics in terms of the torque matrix element than is possible by ``brute-force'' calculation of the dynamical spin-density matrix. 

The Elliott-Yafet mechanism has also been applied to the demagnetization dynamics in ferromagnets.~\cite{Koopmans:PRL2005,Steiauf.2009,Krauss:2009gc,Carva:2011dp,Essert.2011,Koopmans:NatMat2010} None of these approaches avoids all of the three problems above. 

\section{Spin-dependent electron-phonon scattering dynamics\label{e-pn-dynamics}}

In this section we present a derivation of spin-dependent carrier dynamics due to electron-phonon interactions that avoids the three problems listed above. We first derive the interaction hamiltonian between electrons and phonons including the long-range contribution of the Coulomb potential, and specialize to the long-wavelength limit. With this interaction we derive the equation of motion for the spin density matrix including phonon-assisted correlation functions.  We identify the torque matrix element that determines the incoherent dynamics of the average spin. Finally, we derive the scattering limit of the dynamical equation.

\subsection{Electron-phonon interaction hamiltonian}

We start our derivation by writing down the interaction potential of a single electron in a lattice in terms of lattice site coordinates
$\vec R_{n}$ within the rigid-ion approximation as $\hat{v}_{\text{e-L}}(\vec x,\{\vec R_{n}\})=\sum_{n}\hat{v}_{\text{e-ion}}(\vec x-\vec R_{n})$,
where
\begin{equation}
\hat{v}_{\text{e-ion}}(\vec x)=v_{\text{eff}}(\vec x)-\xi\big[\nabla v_{\text{eff}}(\vec x)\times\hat{\vec p}\big]\cdot\hat{\vec s}\ .\label{eq:e-L-Potential_mitQ}
\end{equation}
Here, $v_{\text{eff}}$ is the effective electrostatic Coulomb interaction between the electron and the ionic core and $\xi=\hbar/(4m^{2}c^{2})$. We denote single-particle operators acting on the electronic space and spin variables by small letters and a hat; for instance, $\hat{\vec p}=-i\hbar\nabla$. The second term in \eqref{eq:e-L-Potential_mitQ} is the spin-orbit interaction. We remind the reader that, due to lack of rotational symmetry, this potential does not commute with the electronic angular momentum operator $\hat{\vec x}\times\hat{\vec p}$ of the electrons, and therefore does not conserve the electronic orbital angular momentum. The rigid-ion approximation places no fundamental restriction on the following development, but the equations of motion become more complicated without it.

We follow the usual treatment of the electron-phonon coupling by considering small deviations $\vec R_{n}=\vec R_{n}^{(0)}+\vec Q_{n}$ around
the equilibrium configuration $\vec R_{n}^{(0)}$ and expand $v_{\text{eff}}(\vec x-\vec R_{n})$ with respect to $\vec R_{n}$. The resulting electron-phonon hamiltonian in second quantization is
\begin{equation}
H_{\text{e-pn}}=\sum_{n}\int\Psi^{\dagger}(\vec x)\Big[\vec Q_{n}\cdot\frac{{\partial\hat{v}_{\text{e-ion}}(\vec x-\vec R_{n})}}{\partial\vec R_{n}}\Big|_{\vec R_{n}^{(0)}}\Big]\Psi(\vec x)\ d^{3}x\label{eq:e-lattice-hamiltonian}\, ,
\end{equation}
where $\Psi\equiv(\Psi_{\uparrow},\Psi_{\downarrow})^{T}$ denotes a spinor field operator. Note that $\partial\hat{v}_{\text{e-ion}}/\partial\vec R_{n}$
is an operator in spin space. 

While the properties of the electron-phonon interaction can be discussed without specifying a model for the single-particle band structure, we also need the electronic single-particle contribution to the hamiltonian for the derivation of the equations of motion. We thus consider the model hamiltonian $H=H_{\text{e}} + H_{\text{e-pn}}$ with 
\begin{equation}
	H_{\text{e}} = \int_{\mathcal{V}}\Psi^{\dagger}(\vec x) \big[\frac{\hat{\vec{p}}^2}{2m_0} + \sum_n \hat{v}_{\text{e-ion}}(\vec{x}-\vec{R}_n^{(0)}) \big]\Psi(\vec x)\, d^{3}x\, .
	\label{eq:e-hamiltonian}
\end{equation}

We follow the standard approach~\cite{Mahan} and expand the field operators according to 
\begin{equation}
\Psi(\vec x)=\frac{1}{\sqrt{\mathcal{V}}}\sum_{\mu}\sum_{\vec k}\varphi_{\mu\vec k}(\vec x)c_{\mu\vec k}\, ,
\end{equation}
where 
\begin{equation}
\varphi_{\mu\vec k}(\vec{x}) = e^{i\vec k\cdot\vec x} u_{\mu\vec k}(\vec x)
\end{equation}
are Bloch spinor wavefunctions with band and crystal momentum labels $(\mu\vec k)$. $c_{\mu\vec k}$ is the corresponding destruction operator, and $\mathcal{V}$ is the normalization volume of the crystal. We use the convention that a Greek index includes band and pseudospin index, i.e., $\mu =(b,\lambda)$.

Instead of the Bloch wavefunctions, which are orthogonal on the whole crystal volume~$\mathcal{V}$, we will mainly deal with the lattice periodic $u_{\mu\vec k}$'s, which are defined and orthogonal on the unit cell~$\Omega$, because we intend to derive matrix elements on the unit cell~$\Omega$.  Matrix elements of any \emph{single-electron} operator $\hat{a}$ that occur in the present paper have the meaning 
\begin{equation}
\langle{u_{\mu \vec k}}|\hat{a}u_{\mu \vec{k}'}\rangle_{\Omega}
:=\sum_{\sigma}\frac{1}{\Omega}\int_{\Omega}[u_{\mu\vec k}^{*}(\vec x)]_{\sigma}\big[\hat{a}u_{\mu'\vec k'}(\vec x)\big]_{\sigma}d^{3}x\, .
\label{eq:scalar-product}
\end{equation}

By virtue of Eq.~\eqref{eq:e-L-Potential_mitQ}, the electron-phonon interaction hamiltonian~\eqref{eq:e-lattice-hamiltonian} can be split into two terms
\begin{equation}
H_{\text{e-pn}}=H_{\text{e-pn}}^{\mathrm{El}}+H_{\text{e-pn}}^{\mathrm{Ovhsr}},
\end{equation}
the first one of which we call Elliott and the second Overhauser hamiltonian. The Elliott hamiltonian itself is spin diagonal
\begin{equation}
H_{\text{e-pn}}^{\mathrm{El}}=-\sum_{n}\sum_{\vec{k},\vec{k}'}\sum_{\mu\mu'}\frac{1}{\mathcal{V}}\int \varphi^*_{\mu \vec{k}}(\vec{x}) 
\vec{Q}_n\cdot \boldsymbol{\nabla} v_{\mathrm{eff}} (\vec{x}-\vec{R}_n^{(0)})  \varphi_{\mu' \vec{k}'}\left(\vec{x}\right) d^{3}x\label{eq:e-elliott-hamiltonian},
\end{equation}
whereas the Overhauser hamiltonian contains the electronic spin operator $\hat{\vec s}$ 
\begin{equation}
H_{\text{e-pn}}^{\mathrm{Ovhsr}}=\sum_{n}\sum_{\vec{k},\vec{k'}}\sum_{\mu\mu'}\xi\frac{1}{\mathcal{V}}\int  \varphi^*_{\mu \vec{k}}\left(\vec{x}\right) \left[\boldsymbol{\nabla} \left(\vec{Q}_n\cdot \boldsymbol{\nabla} v_{\mathrm{eff}} (\vec{x}-\vec{R}_n^{(0)})  \right)\times \hat{\vec{p}}\right] \cdot \hat{\vec{s}}  
\; \varphi_{\mu' \vec{k'}}\left(\vec{x}\right) d^{3}x
\label{eq:e-Overhauser-hamiltonian}.
\end{equation}
We next expand the phonon displacement operators according to~\cite{Mahan}
\begin{equation}
\vec{Q}_{n}(t)=i\sum_{\vec q,\lambda}x_{\vec q,\lambda}^{(0)}B_{\vec q,\lambda}e^{i\vec q\cdot\vec R_{n}^{(0)}}\:\boldsymbol{\varepsilon}_{\vec q,\lambda}\, ,\label{eq:PhononApproach}
\end{equation}
with $B_{\vec q,\lambda}\equiv b_{-\vec q,\lambda}^{\dagger}+b_{\vec q,\lambda}$, where $b_{\vec q,\lambda}^{(\dagger)}$ annihilates (creates) phonons with momentum $\vec q$ in the first Brillouin zone and mode index $\lambda$. We use the abbreviation of $x_{\vec q,\lambda}^{(0)}=\sqrt{\hbar/(2\varrho\mathcal{V}\omega_{\vec q,\lambda})}$, which has the unit of length, where $\varrho$ is the density. The phonon dispersion is denoted by $\omega_{\vec q,\lambda}$. Further, $\boldsymbol{\varepsilon}_{\vec q,\lambda}$ is the polarization vector of the phonon mode with the property $(\boldsymbol{\varepsilon}_{\vec q,\lambda})^{*}=-\boldsymbol{\varepsilon}_{-\vec q,\lambda}$. We introduce the Fourier transformation of the interaction potential in the form~\cite{Mahan}
\begin{equation}
v_{\mathrm{eff}}\left(\vec{x}\right)=\frac{1}{N}\sum_{\vec q \in \text{1BZ}}\sum_{\vec{G}}v_{\mathrm{eff}}\left(\vec{q}+\vec{G}\right)e^{i\left(\vec{q}+\vec{G}\right)\cdot\vec{x}}\label{eq:FT Potential},
\end{equation}
which has a long-range part ($\vec{G}=0$) and short-range part ($\vec{G} \neq 0$).

Now we split the integral over the crystal volume $\mathcal{V}$ according to
\begin{equation}
\frac{1}{\mathcal{V}} \int_{\mathcal{V}} f\left(\vec{x}\right)d^3x=\frac{1}{N\Omega}\sum_n\int_{\Omega} f\left(\vec{x}+\vec{R}_n^{(0)}\right)d^3x,
\end{equation}
use the periodicity $u_{\mu\vec{k}}(\vec{x}+\vec{R}^{(0)}_n)=u_{\mu\vec{k}}(\vec{x})$ and the relation $\sum_n e^{i\vec{R}_n^{(0)}\cdot\left(\vec{q}+\vec{k}-\vec{k}'\right)}=N\delta _{\vec{q},\vec{k}-\vec{k}'}$, which neglects Umklapp processes.
We thus obtain for the e-pn coupling hamiltonian~\eqref{eq:e-lattice-hamiltonian}
\begin{equation}
H_{\text{e-pn}}=\sum_{\vec k\vec q}\sum_{\mu\mu'}\sum_{\lambda}g_{\mu\vec k + \vec q,\mu'\vec k}^{(\lambda)}B_{\vec q,\lambda}c_{\mu\vec k + \vec q}^{\dagger}c_{\mu'\vec k}\, ,\label{eq:general_e-ph-Ham}
\end{equation}
with the matrix element 
\begin{equation}
g_{\mu\vec k + \vec q,\mu'\vec k}^{(\lambda)}=\langle u_{\mu\vec k + \vec q}|\hat{v}_{\vec k + \vec q,\vec k}^{(\lambda)}u_{\mu'\vec k}\rangle_{\Omega}
\label{eq:g-matrix}
\end{equation}
in terms of the \emph{e-pn interaction operator} 
\begin{equation}
\hat{v}_{\vec k + \vec q,\vec k}^{(\lambda)} =\sum_{\vec G}x_{\vec q,\lambda}^{(0)}e^{i\vec G\cdot\vec x}v_{\text{eff}}(\vec q+\vec G)\big[\boldsymbol{\varepsilon}_{\vec q, \lambda}\cdot(\vec q+\vec G)\big]\\
\big\{\underbrace{1}_{\text{Elliott}}+ \underbrace{(-i\xi)[(\vec q+\vec G)\times(\hat{\vec p} + \hbar\vec k)]\cdot\hat{\vec s}}_{\text{Overhauser}}\big\}.
\label{eq:v_k}
\end{equation}
The ``1'' in the curly braces in the above expression comes from the Elliott term~\eqref{eq:e-elliott-hamiltonian} and the rest from the Overhauser term~\eqref{eq:e-Overhauser-hamiltonian}. The $\vec q$ and $\vec k$ vectors are restricted to the first Brillouin zone, and we have already neglected Umklapp processes, which contribute if $\vec k+\vec q$ lies outside the first BZ. The long-range part of this expression results from the $\vec G=0$ contribution to the sum, whereas the sum over the $\vec G\neq0$ defines the short-range part of the matrix elements. 
Grimaldi and Fulde~\cite{Grimaldi:1997td} demonstrate that the long-range part of the matrix element is most important in the long-wavelength limit, so that we explicitly isolate the long-range part in the following. 

We will use the e-pn interaction matrix element in the long-wavelength limit as is customary in semiconductor spintronics.~\cite{Wu:2010jf}
Grimaldi and Fulde~\cite{Grimaldi:1997td} also demonstrate how the long-range contribution is screened so that $v_{\text{eff}}(\vec{q})$ has a well defined $\vec q\to0$ (long-wavelength) limit. In this limit, the long-range interaction operator is
\begin{equation}
\hat{v}_{\vec k + \vec q,\vec k}^{(\lambda)}=x_{\vec q,\lambda}^{(0)}v_{\text{eff}}(\vec{q}\to0)\,(\vec{q}\cdot\boldsymbol{\varepsilon}_{\vec q,\lambda})\big\{\underbrace{1}_{\text{Elliott}}-\underbrace{i\xi[\vec q\times(\hat{\vec p}+\hbar\vec k)]\cdot\hat{\vec s}}_{\text{Overhauser}}\big\}\label{eq:m-longrange}.
\end{equation}
Note that for small $q$ the electrons couple exclusively to longitudinal phonons, where the polarization vector $\boldsymbol{\varepsilon}_{\vec q,\lambda}$ points in the same direction as the wave-vector $\vec q$.

\subsection{Equation of motion for spin density-matrix}

In this section we derive the equation of motion for the reduced spin density matrix  
\begin{equation}
\rho_{\vec k}^{\mu\mu'}\equiv\langle\hat{c}_{\mu\vec k}^{\dagger}\hat{c}_{\mu'\vec k}\rangle
\end{equation}
which determines all single-particle properties of the electronic ensemble.

We will for definiteness also include coherent terms and eventually take the scattering limit. Therefore we need the single-particle hamiltonian~\eqref{eq:e-hamiltonian} in diagonal form
\begin{equation}
	H_{\text{e}} = \sum_{\mu} \epsilon_{\mu\vec{k}}c^{\dagger}_{\mu\vec{k}} c_{\mu\vec{k}}.
	\label{eq:h-e-finite}
\end{equation}
Here, we use $\vec{k}\cdot \vec{p}$ theory,~\cite{Ilinskii-Keldysh,Haug-Koch} i.e., the~$u_{\mu\vec k}$'s and energy dispersions $\epsilon_{\mu\vec k}$ are determined as the solution of the eigenvalue problem of the self-adjoint $\vec{k}\cdot \vec{p}$ operator $\hat{h}_{\text{eff}}(\vec{k})$
\begin{equation}
\hat{h}_{\text{eff}}(\vec k)u_{\mu\vec k}=\epsilon_{\mu\vec k}u_{\mu\vec k}
\label{eq:kp-eigenvalue}
\end{equation} 
for lattice-periodic $u_{\mu\vec{k}}$'s on $\Omega$. This guarantees that, for each $\vec{k}$, the eigenvalue spectrum $\epsilon_{\mu\vec{k}}$ is discrete and the $u_{\mu\vec k}$ form a complete set, i.e.,
\begin{equation}
\sum_{\mu}|u_{\mu \vec k}\rangle\langle{u_{\mu \vec k}}|=\mathbb{1}.
\label{eq:u-complete}
\end{equation}
Further, the $u_{\mu\vec{k}}$'s for each $\vec{k}$ are orthogonal with regard to the scalar product~\eqref{eq:scalar-product}. In semiconductors, the $k\cdot p$ hamiltonian operator can be approximated on a subset of bands by an effective hamiltonian, such as the Luttinger hamiltonian.~\cite{Haug-Koch} The number of bands included in~\eqref{eq:h-e-finite} determines the dimension of the reduced density matrix $\rho_{\vec k}$. 

The equation of motion for the electronic spin-density matrix $\rho_{\vec k}^{\mu\mu'}$ due to~\eqref{eq:general_e-ph-Ham} and ~\eqref{eq:h-e-finite} is
\begin{equation}
\frac{\partial}{\partial t} \rho_{\vec{k}}^{\mu\mu'}=\frac{i}{\hbar} \big(\epsilon_{\mu\vec{k}} - \epsilon_{\mu'\vec{k}} \big) 
	\rho_{\vec k}^{\mu\mu'} + \frac{\partial}{\partial t} \rho_{\vec k}^{\mu\mu'}\Big|_{\text{e-pn}}\, ,
	\label{eq:rho-coherent}
\end{equation}
with
\begin{equation}
\begin{split}
\frac{\partial}{\partial t}\rho_{\vec k}^{\mu\mu'}\Big|_{\text{\text{e-pn}}}  &=\frac{i}{\hbar}\sum_{\lambda}\sum_{\nu\vec q}
\big\{  \langle u_{\nu\vec k + \vec q}|\hat{v}_{\vec k + \vec q,\vec k}^{(\lambda)}u_{\mu\vec k}\rangle_{\Omega}
\langle B_{\vec q ,\lambda} c_{\nu\vec k + \vec q}^{\dagger} c_{\mu'\vec k}\rangle \\
&\quad\qquad\qquad- \langle u_{\mu'\vec{k}}|\hat{v}_{\vec k,\vec k + \vec q}^{(\lambda)}u_{\nu\vec k + \vec{q}}\rangle_{\Omega}
\langle B_{-\vec q ,\lambda} c_{\mu\vec k}^{\dagger} c_{\nu\vec k + \vec q}\rangle\big\}.
\label{eq:general_e-ph_EOM}
\end{split}
\end{equation}
Equation~\eqref{eq:rho-coherent} is the full equation of motion that contains the coherent contribution as well as the e-pn interaction contribution~\eqref{eq:general_e-ph_EOM}, in which the phonon-assisted electronic density matrix $\langle b_{\vec q ,\lambda} c_{\nu\vec k + \vec q}^{\dagger} c_{\mu'\vec k}\rangle$  appears.~\cite{RossiKuhn:2002,Haug-Koch} The Elliott-Yafet mechanism arises from the e-pn interaction contribution in~\eqref{eq:rho-coherent}. To study the EY mechanism in isolation, we will in the following neglect the coherent contributions to the equation of motion of the spin density-matrix. When we specialize the results to the case of a pair of Kramers degenerate bands in Sec.~\ref{sec:relaxation-time} below, the coherent contributions vanish exactly.

\subsection{Dynamics of the average spin due to incoherent e-pn interaction\label{sec:average-spin-e-pn}}

Before we make approximations to the phonon-assisted density matrix, we compute the dynamics of the average spin 
\begin{equation}
S_{\alpha} = \Big\langle \int_{\mathcal{V}} \Psi^{\dagger}(\vec x)\hat{s}_{\alpha}\Psi(\vec x)\,d^{3}x\Big\rangle 
                  =\sum_{\vec k}\sum_{\mu\mu'}\langle{u_{\mu \vec k}}|\hat{s}_{\alpha}u_{\mu' \vec k}\rangle_{\Omega}\;\rho_{\vec k}^{\mu\mu'}.
\label{eq:S-definition}
\end{equation}
By combining~\eqref{eq:general_e-ph_EOM} and~\eqref{eq:S-definition} we obtain
\begin{equation}
\begin{split}
\frac{\partial}{\partial t}&S_\alpha \Big|_{\text{\text{e-pn}}}  =\frac{i}{\hbar}\sum_{\lambda}\sum_{\nu\mu\mu'}\sum_{\vec{k},\vec{q}}
\langle{u_{\mu \vec k}}|\hat{s}_{\alpha}u_{\mu' \vec k}\rangle_{\Omega} \\
&\times \big\{ \langle u_{\nu\vec k + \vec q}|\hat{v}_{\vec k + \vec q,\vec k}^{(\lambda)}u_{\mu\vec k}\rangle_{\Omega}
\langle B_{\vec q ,\lambda} c_{\nu\vec k + \vec q}^{\dagger} c_{\mu'\vec k}\rangle
- \langle u_{\mu'\vec{k}}|\hat{v}_{\vec k,\vec k + \vec q}^{(\lambda)}u_{\nu\vec k + \vec{q}}\rangle_{\Omega}
\langle B_{-\vec q ,\lambda} c_{\mu\vec k}^{\dagger} c_{\nu\vec k + \vec q}\rangle\big\}\ .
\end{split}
\label{eq:S-dot}
\end{equation}
We can employ the completeness property~\eqref{eq:u-complete} of the $u_{\mu\vec{k}}$'s to show that this has the form  
\begin{equation}
\frac{\partial}{\partial t}S_{\alpha}\Big|_{\text{e-pn}} =\sum_{\vec k\vec q}\sum_{\mu\mu',\lambda}\langle B_{\vec q,\lambda}c_{\mu'\vec k + \vec q}^{\dagger}c_{\mu\vec k}\rangle\langle{u_{\mu' \vec k + \vec q}}|\big(\hat{\vec t}_{\vec k + \vec q,\vec k}^{(\lambda)}\big)_{\alpha}u_{\mu \vec k}\rangle_{\Omega},
\label{eq:e-ph_SpinChange}
\end{equation}
with the torque vector operator due to the e-pn interaction 
\begin{equation}
\hat{\vec{t}}_{\vec k + \vec q,\vec k}^{(\lambda)}:=\frac{1}{i\hbar}\big[\hat{\vec s},\hat{v}_{\vec k + \vec q,\vec k}^{(\lambda)}\big].
\end{equation}
We will denote the matrix element of this operator by
\begin{equation}
	\vec{t}_{\mu'\vec{k}+\vec{q},\mu\vec{k}}^{(\lambda)} := \langle u_{\mu'\vec k + \vec q}|\hat{\vec t}_{\vec k +\vec q, \vec k}^{(\lambda)} u_{\mu\vec k}\rangle.
\end{equation}
This matrix element completely determines the spin change that occurs in a transition ($\mu\vec{k})\to(\mu'\vec{k}+\vec{q})$. This information is not explicit the matrix element~\eqref{eq:g-matrix}. Using \eqref{eq:v_k}, one finds
\begin{equation}
\hat{\vec{t}}_{\vec k + \vec q,\vec k}^{(\lambda)} =\sum_{\vec G} x_{\vec q,\lambda}^{(0)}e^{i\vec G\cdot\vec x}v_{\text{eff}}(\vec q +\vec G)[\boldsymbol{\varepsilon}_{\vec q, \lambda}\cdot(\vec q +\vec G)]\big\{\underbrace{0}_{\text{Elliott}}+\underbrace{(-i\xi)[(\vec q+\vec G)\times(\hbar\vec k+\hat{\vec p})]\times\hat{\vec s}}_{\text{Overhauser}}\big\}.\label{eq:t_k}
\end{equation}
The expression \eqref{eq:e-ph_SpinChange} together with \eqref{eq:t_k} is already an important result, as it shows that\emph{only} the Overhauser contribution to the e-pn interaction gives rise to a non-vanishing torque \eqref{eq:t_k}. The Elliott contribution is spin diagonal and therefore its contribution to the torque vanishes. We stress that this result is general enough to include materials with Kramers degeneracy (and ``spin hot spots'') as well as non-spin degenerate systems with avoided level crossings due to spin-orbit coupling. In all cases, the Elliott contribution to the torque matrix element \emph{does not change the average spin}. This conclusion does not depend on the single-particle basis $u_{\mu\vec{k}}$ in the sense that one could apply a $\vec{k}$-dependent unitary transformation to the $u_{\mu\vec{k}}$'s and the $c_{\mu\vec{k}}$'s in~\eqref{eq:e-ph_SpinChange} and~\eqref{eq:t_k}. This argument shows that it is not the spin mixing \emph{in the wave functions} that determines the Elliott torque, and---in this restricted sense---there is no Elliott contribution to spin dynamics in any basis. 

The torque operator $\hat{\vec{t}}_{\vec k + \vec q,\vec k}^{(\lambda)}$, which is due to the Overhauser contribution only, has the long-wavelength limit
\begin{equation}
\hat{\vec{t}}_{\vec k + \vec q,\vec k}^{(\lambda)} =-i\xi x_{\vec q,\lambda}^{(0)}v_{\text{eff}}(\vec{q}\to 0)\, (\vec{q}\cdot\boldsymbol{\varepsilon}_{\vec q, \lambda})[\vec q\times(\hbar\vec k+\hat{\vec p})]\times\hat{\vec s}. \label{eq:t-longrange}
\end{equation}
Finally, we note also that~\eqref{eq:e-ph_SpinChange} and the definition of the torque matrix element is a rather general result for the spin change due to any incoherent electron-boson scattering, for instance, with photons or magnons. 

\subsection{Dynamics of the average spin in the scattering limit}

To compare the results of the present approach with the conventional EY analysis, we specialize the dynamical equation~\eqref{eq:e-ph_SpinChange} for $S_{\alpha}$ for the case of incoherent scattering. To this end, we need to evaluate the phonon-assisted density matrix~$\langle b_{\vec q,\lambda}c_{\mu\vec k+\vec q}^{\dagger}c_{\mu'\vec k}\rangle$.
We do this by truncating its equation of motion at the scattering level to obtain
\begin{equation}
\begin{split}
\frac{\partial}{\partial t}\langle b_{\vec q,\lambda}c_{\mu\vec k+\vec q}^{\dagger}c_{\mu'\vec k}\rangle =&-\frac{i}{\hbar} \left(\epsilon_{\mu'\vec{k}}-\epsilon_{\mu\vec{k}+\vec{q}}+\hbar\omega_{\vec{q},\lambda}-i\hbar\gamma\right)\langle b_{\vec q,\lambda}c_{\mu\vec k+\vec q}^{\dagger}c_{\mu'\vec k}\rangle \\
 &\mbox{}-\frac{i}{\hbar}\sum_{\tau\tau'} g_{\tau'\vec{k},\tau\vec{k} + \vec{q}}\Big\{[\delta_{\mu'\tau'}-\rho^{\tau'\mu'}_{\vec k}]\rho^{\mu\tau}_{\vec k+\vec q}(1+N_{\vec q, \lambda})-[\delta_{\tau\mu}-\rho^{\mu\tau}_{\vec k+\vec q}]\rho^{\tau'\mu'}_{\vec k}N_{ \vec q, \lambda}\Big\},
\end{split}
\end{equation}
where $N_{\vec q,\lambda}=b(\hbar\omega_{\vec q,\lambda}$) are phonon occupation numbers, and performing a Markov approximation, see, e.g.,~Refs.~\onlinecite{RossiKuhn:2002,Haug-Koch}. 

The equation of motion for the phonon-assisted density matrix and, consequently, for the reduced density-matrix $\rho_{\vec k}^{\mu\mu'}$ is determined by the full matrix element including both the Elliott and Overhauser contributions, and one finds for the ensemble averaged spin 
\begin{widetext}
\begin{equation}
\begin{split}\frac{\partial}{\partial t}\vec S & =2\text{Re}\sum_{\vec k \vec q}\sum_{\mu\mu'}\sum_{\lambda}
	\vec{t}_{\mu'\vec{k}+\vec{q},\mu\vec{k}}^{(\lambda)} g_{\tau'\vec{k},\tau\vec{k} + \vec{q}}^{(\lambda)}\frac{1}{(\epsilon_{\mu'\vec{k}+\vec{q}}-\epsilon_{\mu \vec{k}}-\hbar\omega_{\vec q, \lambda})+i\hbar\gamma}\\
 & \qquad\quad\times\sum_{\tau\tau'}\Big\{[\delta_{\mu\tau'}-\rho^{\tau'\mu}_{\vec k}]\rho^{\mu'\tau}_{\vec k+\vec q}(1+N_{\vec q, \lambda})-[\delta_{\tau\mu'}-\rho^{\mu'\tau}_{\vec k+\vec q}]\rho^{\tau'\mu}_{\vec k}N_{ \vec q, \lambda}\Big\}.
\end{split}
\label{dS-dt-1}
\end{equation}
\end{widetext}
We will assume that the phonon system is not changed appreciably by the electronic dynamics so that the $N_{\vec q,\lambda}=b(\hbar\omega_{\vec q,\lambda}$) are the \emph{equilibrium} phonon occupations given by the Bose function~$b\left(\hbar\omega_{\vec q,\lambda}\right)$. At this level of generality, we still have to deal with the reduced density matrix on the right-hand side.

\section{Relaxation time for Kramers degenerate bands\label{sec:relaxation-time}}

We now specialize to the case of a pair of Kramers degenerate bands with a quasi-equilibrium spin-expectation value to obtain a relation for the spin relaxation, which generalizes Yafet's treatment. 

\subsection{Eigenstates for Kramers degenerate bands}

Kramers degenerate states including spin-orbit coupling can be assumed to have the form~\cite{Fabian:1998tb,Fabian:1999tv,Cheng:2010js}
\begin{align}
u_{\vec{k}\Uparrow}(\vec{x}) &= a_{\vec{k}}(\vec{x})|\uparrow\rangle +b_{\vec{k}}(\vec{x})|\downarrow\rangle \label{u-uparrow}, \\
u_{\vec{k}\Downarrow} (\vec{x}) &=a_{-\vec{k}}^{*}(\vec{x})|\downarrow\rangle -b^*_{-\vec{k}}(\vec{x})|\uparrow\rangle \label{u-downarrow}.
\end{align}
Part of this assumption is that these states are labeled by pseudo spin indices and that, generally, $\frac{1}{\Omega}\int_\Omega |b_{\vec{k}}(\vec{x})|^2\,d^3x$ is much smaller than $\frac{1}{\Omega}\int_\Omega |a_{\vec{k}}(\vec{x})|^2\,d^3x$. Because of the degeneracy, there is an ambiguity to the definition of these states, as any superposition will also be an eigenstate. In accordance with Refs.~\onlinecite{Fabian:1998tb,Fabian:1999tv,Cheng:2010js} we choose them to fulfill
\begin{equation}
\label{eq:s-offdiagonal}
 \langle u_{\Uparrow,\vec{k}}|\hat{s}_z u_{\Downarrow,\vec{k}}\rangle = 0 .
 \end{equation}
This is equivalent to the statement that the chosen states $u_{\Uparrow\vec{k}}$, $u_{\Downarrow\vec{k}}$ diagonalize $\hat{s}_z$ in the degenerate subspace. The choice is important, if one wants to attach some importance to the magnitude of $\frac{1}{\Omega}\int_\Omega |b_{\vec{k}}(\vec{x})|^2\,d^3x$ as spin mixing parameter. In ab-initio calculations one often uses a small external magnetic field to enforce a quantization direction so that~\eqref{eq:s-offdiagonal} is fulfilled.

The condition~\eqref{eq:s-offdiagonal} does not imply that the electrons are in pure spin-up or spin-down states characterized by the same two-dimensional spinors at each $\vec{k}$. In fact, we have
\begin{equation}
 u_{\Uparrow,\vec{k}}\neq |\uparrow\rangle \qquad \text{and} \qquad u_{\Downarrow,\vec{k}}\neq |\downarrow\rangle,
\end{equation}
as well as $|\langle u_{\lambda\vec{k}}|\hat{s}_z u_{\lambda\vec{k}}\rangle|\leq\hbar/2$ for $\lambda = \Uparrow$, $\Downarrow$. For more details, see appendix~\ref{sec:spin-expectation}.

\subsection{Quasi-equilibrium with spin polarization}

In order to derive a characteristic rate for the relaxation of a small excess spin polarization~$\delta S_{z}$ at a temperature $T$, we first need to determine the reduced density matrix for this case. We assume the system to be in a quasi-equilibrium with a given spin-expectation value~$\delta S_{z}$.  Thus we must determine the quasi-equilibrium distribution for the eigenenergies of  the generalized grand-canonical $\vec k\cdot \vec p$ hamiltonian 
\begin{equation}
\hat{\mathfrak{k}}(\vec k)\equiv\hat{h}_{\text{eff}}(\vec k)-\mu-\zeta_{z}\hat{s}_{z}\ .
\end{equation}
Here, $\zeta_{z}$, the spin accumulation, is the Lagrange parameter needed to obtain the finite spin expectation value. Since $\hat{h}_{\text{eff}}(\vec k)$ includes spin-orbit coupling it does not commute with $\hat{s}_{z}$, so that the eigenstates and eigenenergies of 
\begin{equation}
\hat{\mathfrak{k}}(\vec k)w_{\mu\vec k}=E_{\mu\vec k}w_{\mu\vec k}
\end{equation}
are, in general, \emph{not} identical to the $u_{\mu\vec k}$'s that result as eigenfunctions of $\hat{h}_{\mathrm{eff}}(\vec k)$, cf.~Eq.~\eqref{eq:kp-eigenvalue}. 

We assume that $\zeta_{z}$ is small and determine $E_{\mu\vec{k}}$ perturbatively in the degenerate subspace of the $u_{\mu \vec{k}}$'s. Since we chose the $u_{\mu\vec{k}}$'s that diagonalize $\hat{s}_z$ in the degenerate subspace, cf.~\eqref{eq:s-offdiagonal}, this gives 
\begin{equation}
E_{\mu\vec k}\simeq\epsilon_{\vec k}-\mu-\zeta_{z}\langle u_{\mu\vec k}|\hat{s}_{z} u_{\mu\vec k}\rangle \ . \label{eq: GK energies}
\end{equation}
If we also assume that other bands are sufficiently far away, the eigenvectors are $w_{\mu \vec k} \simeq u_{\mu \vec k}$. If there are other bands close to the bands for which the spin relaxation is computed, one needs to include an eigenvector correction for $w_{\mu \vec k}$.  

The quasi-equilibrium reduced density matrix in the eigenbasis of the grand canonical hamiltonian is diagonal and depends only on the grand-canonical energies~$E_{\mu \vec k}$, i.e., 
\begin{equation}
\rho_{\vec{k}}^{\mu\mu'}=\delta_{\mu,\mu'} f(E_{\mu\vec{k}})\ ,
\end{equation}
where $f(\epsilon)=[\exp(\beta \epsilon)+1]^{-1}$ is the Fermi function. Since $\zeta_z$ is small, we keep only the first order result in $\zeta_z$ and with~\eqref{eq: GK energies} we find for the quasi-equilibrium distribution
\begin{equation}
	\rho^{\mu\mu'}_{\vec k}= \delta_{\mu,\mu'}\big[f(\epsilon_{\vec k}-\mu) 
		+ \Delta_{\vec k} \zeta_{z} \langle u_{\mu\vec k}|\hat{s}_{z}u_{\mu\vec k}\rangle\big],
\label{rho-GK}
\end{equation}
where we have defined $\Delta_{\vec k} \equiv -\frac{df}{d\epsilon}|_{\epsilon_{\vec k}-\mu}$. For low temperatures, this function approaches a $\delta$-function concentrated at $\epsilon_{\vec k}-\mu$. From the reduced density matrix~\eqref{rho-GK} we find the relation $\zeta_{z}=\delta S_{z}/\mathcal{N}$ between $\delta S_z$ and $\zeta_z$ with the normalization 
\begin{equation}
\mathcal{N}=\sum_{\mu\vec k}\langle u_{\mu\vec k}|\hat{s}_{z} u_{\mu\vec k}\rangle^{2}\Delta_{\vec k}\ .
\end{equation}

\subsection{EY relaxation time}
 
To obtain the close-to-equilibrium spin relaxation rate defined by 
\begin{equation}
\frac{\partial}{\partial t}\delta S_{z} =-\frac{\delta S_{z}}{\tau_{\text{SR}}},
\end{equation}
 we insert~\eqref{rho-GK} in~\eqref{dS-dt-1} and linearize. Then, as in Yafet's original treatment and its extension by F\"{a}hnle and coworkers,~\cite{Steiauf:2010ea} $\zeta_{z}$ drops out and we find
\begin{widetext}
\begin{equation}
\begin{split}
\frac{1}{\tau_{\text{SR}}} 
	& =-\frac{2}{\mathcal{N}}\text{Re}\sum_{\vec k \vec q}\sum_{\mu\mu'}\sum_{\lambda}
		\big(t_{\mu'\vec k + \vec q, \mu\vec k}^{(\lambda)}\big)_{z} 
		g_{\mu\vec k, \mu'\vec k +\vec q}^{(\lambda)}  
		\langle u_{\mu\vec k + \vec q}|\hat{s}_{z} u_{\mu\vec k + \vec q}\rangle\Delta_{\vec k + \vec q}\\
	 & \qquad\qquad\qquad\times\Big\{\frac{N_{\vec q,\lambda}+f(\epsilon_{\mu\vec k}-\mu)}{\epsilon_{\mu' \vec k + \vec q}-\epsilon_{\mu\vec k}+\hbar\omega_{\vec q,\lambda}+i\hbar\gamma}
	 	+\frac{1+N_{\vec q,\lambda}-f(\epsilon_{\mu\vec{k}}-\mu)}{\epsilon_{\mu' \vec k + \vec q}-\epsilon_{\mu\vec k}-\hbar\omega_{\vec{q},\lambda}+i\hbar\gamma}\Big\}\ .
\end{split}
\label{eq:EY-new}
\end{equation}
\end{widetext}
This expression for the spin relaxation, or $T_1$, time is a more physically transparent result compared to Yafet's because it relates the spin dynamics directly to a torque matrix element. It improves on the original Yafet result by including the correct bookkeeping for spin instead of accounting for pseudo-spin flips. Instead of $|g^{(\lambda)}_{\mu \vec k,\mu'\vec k'}|^2$, i.e., the squared modulus of the e-pn matrix element, as it occurs in Yafet's result, here matrix elements of the torque, the electron-phonon interaction and the spin appear. As shown in Sec.~\ref{sec:average-spin-e-pn} the torque matrix element only has an Overhauser contribution, whereas in the $g$ matrix element both Elliott and Overhauser terms contribute. Regarding the $g$ matrix element in the long wavelength limit, Yafet~\cite{Yafet:1963iw} and numerical evaluations~\cite{Cheng:2010js} of his spin-relaxation result find a cancellation between Elliott and Overhauser contributions in the short range contribution, whereas Grimaldi and Fulde~\cite{Grimaldi:1997td} find for the long-wavelength limit a dominating long-range Elliott contribution to $g$. 

Our result~\eqref{eq:EY-new} is not limited to small spin mixing and is qualitatively different from the original Yafet formula, as there is no pure Elliott contribution.  It will make it easier to compute accurate EY spin relaxation times numerically and to compare different contributions to spin dynamics by computing the relevant combination of torque and $g$ matrix elements.
 

\section{Conclusion\label{sec:conclusion}}

In conclusion, we have presented an investigation of the carrier-spin dynamics due to incoherent electron-phonon scattering in the presence of spin-orbit coupling, as originally considered by Overhauser, Elliott, and Yafet. We examined the dynamical equation for the reduced density matrix in terms of the phonon assisted density matrix including the contributions of the spin-diagonal (electrostatic) electron-phonon interaction as well as the phonon-modulated spin-orbit coupling. We showed that the central quantity to account for the spin change is a torque matrix element, which gives a physically more appealing picture of spin dynamics due to the EY mechanism. Our results show that the electron-phonon ``Elliott contribution,'' i.e., the spin mixing in the Bloch states, has no impact on the torque matrix element due to the electron-phonon interaction. This important property of spin-dependent carrier dynamics gets lost if one works only with pseudo-spin distributions because these do not characterize the spin completely. Only the explicitly spin-dependent phonon-induced modulation of the spin-orbit coupling as introduced by Overhauser gives rise to a non-vanishing torque due to an incoherent scattering mechanism. Finally, we presented an explicit expression for the spin relaxation, or $T_{1}$, time close to equilibrium, which is valid for arbitrary spin mixing.

\begin{acknowledgments}
Svenja Vollmar has been supported by the Excellence Initiative (DFG/GSC 266). Alexander Baral and Svenja Vollmar contributed equally to this work. We thank D. H\"{a}gele (Bochum) for helpful discussions.
\end{acknowledgments}

\appendix

\section{Spin expectation values for degenerate bands\label{sec:spin-expectation}} 

We show here in detail that, for the case of degenerate bands, the electron spin operator can be diagonalized in the space of the degenerate bands, even if spin-mixing due to spin-orbit coupling is present. This is not a new result, but it is important to keep this in mind when dealing with Kramers degenerate wave functions. 

We assume we have arbitrary $u_{\mu\vec{k}}$'s of the general form
\begin{align}
u_{1\vec{k}}(\vec{x}) &= a_{\vec{k}}(\vec{x})|\uparrow\rangle +b_{\vec{k}}(\vec{x})|\downarrow\rangle \label{eq:u1}, \\
u_{2\vec{k}} (\vec{x}) &=a_{-\vec{k}}^{*}(\vec{x})|\downarrow\rangle -b^*_{-\vec{k}}(\vec{x})|\uparrow\rangle, \label{eq:u2}
\end{align}
which are Kramers degenerate. In this subspace, for fixed $\vec{k}$ the matrix of the spin operator $\hat{s}_{z}$ is
\begin{equation}
	s_z  = \frac{\hbar}{2}\begin{pmatrix}
            d_{\vec{k}}  & c _{\vec{k}}\\  
     c^*_{\vec{k}}    & - d_{-\vec{k}}
	\end{pmatrix}
\end{equation}
with the real diagonal elements
\begin{equation}
d_\vec{k}:=\left\langle u_{\vec{k}1}\left|\hat{s}_{z}\right|u_{\vec{k}1}\right\rangle =\int_{\Omega}\frac{d^3x}{\Omega}\left[\left|a_{\vec{k}}\left(\vec{r}\right)\right|^{2}-\left|b_{\vec{k}}\left(\vec{r}\right)\right|^{2}\right],
\end{equation}
and complex off-diagonal elements
\begin{equation}
c_{\vec{k}}:=\left\langle u_{\vec{k}2}\left|\hat{s}_{z}\right|u_{\vec{k}1}\right\rangle =-\int_{\Omega}\frac{d^3x}{\Omega}\left[a_{\vec{k}}\left(\vec{r}\right)b_{-\vec{k}}\left(\vec{r}\right)+a_{-\vec{k}}\left(\vec{r}\right)b_{\vec{k}}\left(\vec{r}\right)\right].
\end{equation}

If $c_{\vec{k}} \neq 0$, diagonalization leads to the eigenstates $u_{\lambda \vec{k}}$ with $\lambda = \Uparrow$, $\Downarrow$
\begin{align}
u_{\Uparrow\vec{k}}(\vec{x}) &= \frac{c_\vec{k}}{L}^*|\uparrow\rangle -\frac{d_{\vec{k}}+d_{-\vec{k}}}{2L}\left[1-\sqrt{1+\frac{4\left|c_\vec{k}\right|^2}{\left(d_{\vec{k}}+d_{-\vec{k}}\right)^2}}\right]|\downarrow\rangle,  \label{eq: eigenstate1} \\
u_{\Downarrow\vec{k}}(\vec{x}) &= \frac{c_\vec{k}}{L}|\downarrow\rangle +\frac{d_{\vec{k}}+d_{-\vec{k}}}{2L}\left[1-\sqrt{1+\frac{4\left|c_\vec{k}\right|^2}{\left(d_{\vec{k}}+d_{-\vec{k}}\right)^2}}\right]|\uparrow\rangle \label{eq: eigenstate2},
\end{align}
where $L$ normalizes the $u_{\lambda \vec{k}}$'s.
The corresponding spin eigenvalues are
\begin{equation}
\frac{\hbar}{2}\times \Big[\frac{1}{2}(d_{\vec{k}}-d_{-\vec{k}})\pm\sqrt{\frac{1}{4}(d_{\vec{k}}+d_{-\vec{k}})^2+|c_\vec{k}|^2}\Big].
\end{equation}
In the presence of inversion symmetry, as in aluminium and silicon, these expressions simplify further. Then $d_{\vec{k}}=d_{-\vec{k}}$ and the eigenvalues are $\pm (\hbar/2) \sqrt{d_{\vec{k}}^2 + |c_{\vec{k}}|^2}$ for $\lambda = \Uparrow$, $\Downarrow$, respectively. This result shows that pseudo spin labels are justified as these states generally have an expectation value of $\hat{s}_z$ of modulus smaller than (or almost equal to) $\pm \hbar/2$. At the same time, the pseudo-spin states are not just rotated pure spin states, which would lead to a spin projection of exactly $\pm\hbar/2$ along the quantization axis. Further, these states may have nonvanishing $\hat{s}_{x}$ and $\hat{s}_{y}$ expectation values.

\bibliographystyle{apsrev4-1}

\end{document}